\documentclass[a4paper,11pt]{article}
\usepackage{amsmath,amsfonts,amssymb,amsthm,amstext,amscd}
\usepackage{latexsym}
\usepackage[hidelinks=true]{hyperref}
\usepackage{graphicx}
\usepackage{caption}
\usepackage{comment}
\usepackage{graphicx}
\usepackage{cite}
\usepackage{color}
\usepackage{setspace}
\usepackage{float}
\usepackage[mathscr]{eucal}
\usepackage{xcolor}
\usepackage{titletoc}
\dottedcontents{section}[1.2em]{}{1em}{1pc}

\def\a{\alpha}

\def\d{\delta}
\def\e{\epsilon}

\def\m{\mu}
\def\n{\nu}

\def\r{\rho}

\def\P{\Psi}
\def\L{\Lambda}

\def\be{\begin{equation}}
\def\ee{\end{equation}}
\def\arr{\begin{array}{rll}}
\def\ea{\end{array}}
\def\bea{\begin{eqnarray}}
\def\eea{\end{eqnarray}}
\newcommand{\mr}[1]{\mathrm{#1}}

\definecolor{darkred}{rgb}{0.9,0.05,0.05}
\definecolor{darkblue}{rgb}{0.05,0.05,0.6}
\definecolor{darkgreen}{rgb}{0.05,0.6,0.05}
\definecolor{brightgreen}{rgb}{0.1,0.9,0.1}

\renewcommand*{\eqref}[1]{%
	\begingroup
	\hypersetup{
		linkcolor=darkblue,
		linkbordercolor=darkblue,
	}%
	\textcolor{darkblue}{(\ref{#1})}%
	\endgroup
}
\hypersetup{linkcolor=red,citecolor=darkgreen,urlcolor=darkred,colorlinks=true}

\begin{document}

\setlength{\skip\footins}{0.8cm}

\begin{titlepage}

\begin{flushright}\vspace*{-1cm}
{\small
%{\tt arXiv:yymm.nnnn} \\
IPM/P-2017/049}\end{flushright}
\vspace{0.5cm}

\begin{center}
\renewcommand{\baselinestretch}{1.5}  %Line spacing
\setstretch{1.5}

{\fontsize{17pt}{20pt}\bf{Cosmological constant is a conserved charge}}
 
\vspace{9mm}
\renewcommand{\baselinestretch}{1}  %Line spacing
\setstretch{1}

\centerline{\large{Dmitry Chernyavsky$^\dagger$\footnote{e-mail: chernyavsky@tpu.ru} and  Kamal Hajian$^\ast$\footnote{e-mail: kamalhajian@ipm.ir}}}

\vspace{3mm}
\normalsize
\textit{$^\dagger$School of Physics, Tomsk Polytechnic University, 634050 Tomsk, Lenin Ave. 30, Russian Federation}\\
 \textit{$^\ast$School of Physics, Institute for Research in Fundamental
Sciences (IPM), \\P.O. Box 19395-5531, Tehran, Iran}\\
\vspace{5mm}

\begin{abstract}
Cosmological constant can always be considered as the on-shell value of a top form in gravitational theories. The top form is field strength of a gauge field, and the theory enjoys a gauge symmetry. We show that cosmological constant is the charge of global part of the gauge symmetry, and is conserved irrespective of the dynamics of the metric and other fields. In addition, we introduce its conjugate chemical potential, and prove the generalized first law of thermodynamics which includes variation of cosmological constant as a conserved charge. We discuss how our new term in the first law is related to the volume-pressure term. In parallel with the seminal Wald entropy, this analysis suggests that pressure can also be considered as a conserved charge.
\end{abstract}
\end{center}

\let\newpage\relax
\end{titlepage}

\section{Introduction}
One hundred years after introduction of the cosmological constant $\Lambda$ \cite{Einstein:1917}, astronomical observations \cite{Perlmutter:1998np,Riess:1998cb}, as well as AdS/CFT correspondence \cite{Maldacena:1997re,Brown:1986nw}, have put it in the heart of recent researches in gravitational theories.  One of the active lines of research concerning this mysterious constant of nature is studying its thermodynamic effects  which resembles pressure in black hole solutions to gravity \cite{Henneaux:1984ji,Henneaux:1989zc,Teitelboim:1985dp,Caldarelli:1999xj,Sekiwa:2006qj,Cvetic:2010jb,Dolan:2010ha,Dolan:2011xt,Mann_Kubiznak} and contributes to their first law of thermodynamics as a $V\delta P$ term (find some proofs in \cite{Kastor:2009wy,Couch,Hyun:2017nkb}). Nonetheless, there is a shortcoming for this contribution, as well as some unexpected features.  The problem with this term is that $V$ has not a clear-cut definition as a volume associated to the black hole. Unexpected features of this term are I) in order to define the volume $V$, one needs to know the metric inside the black hole, which is in contrast with other coefficients in the first law (e.g. Hawking temperature, horizon angular velocity etc.) which can be found by the information just around the horizon, II) the order of extensive and intensive quantities are reversed; $P$ is expected to be intensive while $V$ to be extensive, which is not the same order of other terms in the black holes first law, III) $P$ is a parameter in the Lagrangian (the $\Lambda$). So, its variation is conceptually different from other variations (e.g. variation of energy, entropy, angular momentum, etc.) which are variations of parameters in the solution.     

In this paper we propose an alternative intensive quantity for the $V$ which is defined using the information on the horizon, and independent of the first law. Moreover, we promote the cosmological constant to be a conserved charge and hence an extensive quantity. This new perspective resolves the issues mentioned above and sheds light on how to find ``$V$" directly from the solution. To this end, we will work in an established but not appreciated  context where $\Lambda$ is considered as the on-shell value (and therefore, a parameter in the \emph{solution}) of a (non)-dynamical top form \cite{Aurilia:1980xj, Duff:1980qv, Hawking:1984hk,Bousso:2000xa}. 

In what follows, first we review how $\Lambda$ can originate from the on-shell value of a top form $\boldsymbol{F}$. Then in the next section, we elaborate the global part of the gauge symmetry of the gauge field in this theory. Calculating the conserved charge of this symmetry (denoted by $C$), we show that $\Lambda$ is proportional to $C^2$.  In the last section, we propose the conjugate chemical potential $\mathit{\Theta}$ for this conserved charge, and prove the generalized first law which enjoys an additional term $\mathit{\Theta}{\d}C$. After exemplifying the Kerr-(A)dS black hole, we discuss how $\mathit{\Theta}{\d} C$ is related to the pressure term $V{\d}P$ in the first law.\\

\noindent\textbf{Review: cosmological constant from a top form}\\
\noindent\textit{Notation:} $[\mu_1\mu_2\dots\mu_p]$ will be used to denote anti-symmetrization of the set of indices $\{\mu_1,\mu_2,\dots,\mu_p\}$  normalized by the factor $\frac{1}{p!}$. Moreover, the standard notation for exterior derivative of a $p$-form $\boldsymbol{a}=\frac{1}{p!}a_{\r_1\dots \r_p}dx^{\r_1}\wedge\dots\wedge dx^{\r_p}$ will be used:
\begin{equation}
d\boldsymbol{a}=(p+1)\,\partial_{[\mu_1}a_{\mu_2\dots\mu_{p+1}]}\,dx^{\mu_1}\wedge\dots\wedge dx^{\mu_{p+1}}.\nonumber
\end{equation}
Here, we review how the cosmological constant $\Lambda$ can be achieved from the introduction of a top form to the Lagrangian. For clarity, we choose the gravity to be the Einstein general relativity, $\mathcal{L}_{_E}=\frac{1}{16{\pi}G}R$. Nevertheless, the construction (and our analysis in the next parts of the paper) is essentially quite general and independent of this choice. In $D$-dimensional space-time, one can introduce a $D$-form $\boldsymbol{F}$  as the exterior derivative of a $(D-1)$-form $\boldsymbol{A}$, i.e. $\boldsymbol{F}=d\boldsymbol{A}$. Accordingly, the action can be modified by the standard kinetic term for this new gauge field
\be\label{Action}
I=\int d^Dx \,(\mathcal{L}_{_E}\pm\mathcal{L}_{_F})=\frac{1}{16 \pi G}\int d^Dx \Big(R\mp\frac{2}{D!} F^2\Big).
\ee
$d^Dx$ is the volume form $\frac{\sqrt{-g}}{D!}\epsilon_{\mu_1\dots\mu_D}dx^{\mu_1}\wedge\dots\wedge dx^{\mu_D}$ with the convention $\epsilon_{01\dots D-1}=+1$ for the Levi-Civita tensor. The gauge field kinetic term is explicitly $F^2\!=\!F_{\mu_1\dots\mu_D}F^{\mu_1\dots\mu_D}$. It is easy to find equations of motion as
\begin{align}\label{EH equation}
R_{\mu\nu}-\frac{1}{2} R g_{\mu\nu}=8\pi G T_{\mu\nu}, \qquad \qquad \nabla_\mu F^{\mu\mu_2\dots\mu_{D}}=0,
\end{align}
in which the energy-momentum tensor $T_{\mu\nu}$ is
\begin{equation}
\hspace*{-0.1cm}T_{\mu\nu}=\frac{\pm 1}{4\pi G(D-1)!} \Big(F_{\mu\r_2\dots\r_D}F_{\nu}{}^{\r_2\dots\r_D}-\frac{1}{2D} F^2 g_{\m\n}\Big).
\end{equation}
The second set of equations in \eqref{EH equation} implies that
\be\label{F_Solution}
F_{\mu_1\dots\mu_D}=c\sqrt{-g}\,\e_{\mu_1\dots\mu_D},
\ee
where we assume positivity of the constant $c$ for later convenience. Substituting this solution into the first set of equations in \eqref{EH equation} and using the identities
$
\epsilon_{\mu_1\dots\mu_D}\epsilon^{\mu_1\dots\mu_D}=\frac{D!}{g}$ and $\epsilon_{\mu\r_2\dots\r_D}\epsilon_{\nu}{}^{\r_2\dots\r_D}=\frac{(D-1)!}{g}g_{\mu\nu}$,
 one finds Einstein equations with a cosmological constant
\begin{equation}\label{Lambda_c dynamic}
R_{\mu\nu}-\frac{1}{2} R g_{\mu\nu}+\Lambda g_{\mu\nu}=0, \qquad\qquad  \Lambda =\pm c^2.
\end{equation}
This result shows that, as far as the equations of motions are concerned, the cosmological constant can be introduced from the on-shell top form $\boldsymbol{F}$.  Note that in this approach parameter $\L$ can be  positive or negative depending on the choice of sign for the kinetic term in \eqref{Action}. Hence, to change the sign of the cosmological constant, one needs to change the sign of the kinetic term $F^2$ in the action \eqref{Action}. Paying attention to \eqref{Lambda_c dynamic}, upper and lower signs correspond to positive and negative $\Lambda$ respectively, and this (upper/lower sign vs. $+$/$-\,\,\Lambda$) will be followed.

Not surprisingly, putting the on-shell field strength $\boldsymbol{F}$ into the Lagrangian \eqref{Action} does \emph{not} reproduce the Lagrangian of Einstein-$\Lambda$ theory of gravity \cite{Duff} (actually it produces Einstein-$\Lambda$ theory, but with a wrong sign for $\Lambda$). To reproduce the Lagrangian as well as the equations of motion, it is possible to add a boundary term to the action \cite{Aurilia:1980xj,Duncan:1989ug,Wu}
\begin{align}\label{Action_Shifted}
I=\int \frac{d^Dx}{16 \pi G} \Big(R\mp\frac{2}{D!} F^2\pm\frac{4 \nabla_\mu \left(A_{\mu_2\dots\mu_D} F^{\mu\mu_2\dots\mu_D}\right)}{(D-1)!}\Big).
\end{align}
This additional term is a surface term and does not change the equations of motion \eqref{EH equation}.  In this construction, the gauge field $A_{\mu_1\dots\mu_{D-1}}$ will be (classically and locally) a non-dynamical field, and its equations of motion will be non-dynamical constraints (find out more on counting degrees of freedom of a $p$-form in \cite{Afshar:2018apx}). Imposing these constraints (putting the solution \eqref{F_Solution} into the Lagrangian \eqref{Action_Shifted}), the Einstein-$\Lambda$ gravity is achieved
\be\label{ActionCosmological-}
I\to \frac{1}{16 \pi G}\int d^Dx \left(R-2 \Lambda \right), \qquad\qquad  \Lambda= \pm c^2.
\ee

Fortunately, our analysis in this paper is insensitive whether one chooses the action \eqref{Action} or  \eqref{Action_Shifted}, as will be explained. Hence, we will use the notation $I$ to refer to anyone of these actions.

\section{$\mathbf{\Lambda}$ is conserved charge of global part of the gauge symmetry}
By construction, $\boldsymbol{F}=d\boldsymbol{A}$, the gauge field  in the action $I$ enjoys the gauge freedom as the symmetry of the theory
\begin{equation}
\boldsymbol{A}\to\boldsymbol{A}+d\boldsymbol{\lambda},\qquad\qquad  \boldsymbol{\lambda}=\frac{\lambda_{\mu_1\dots\mu_{D-2}}}{(D-2)!}\,\,dx^{\mu_1}\wedge\dots\wedge dx^{\mu_{D-2}}.
\end{equation}
In this section, we elaborate the following proposition.
\begin{center}
\emph{Cosmological constant $\Lambda$ is conserved charge associated with the global part of this gauge symmetry}.
\end{center}
We will denote this charge by $C$. To this end, first we set up our tools for calculating conserved charges associated with the symmetries of the action $I$. Then, after introducing the ``global part of the gauge symmetry", we calculate its conserved charge $C$ and show that $\Lambda\propto \pm C^2$.

One can use different standard methods for calculating charge of a given symmetry (as e.g. in \cite{Mann}). Specially, the standard Noether procedure suffices for the analysis here. {Therefore, one might skip this section by taking the definition of the ``generators of global part" (presented below) and jumping to the equation \eqref{Standard Noether} which is the famous Noether charge.} However in order to provide necessary tools for the next section, the formulation we use will be the Lee-Wald covariant formulation of phase spaces \cite{Lee:1990gr,Wald:1993nt,Iyer:1994ys,Wald:1999wa} (initiated and followed in \cite{Ashtekar:1987hia,Ashtekar:1990gc,Crnkovic:1987at,Barnich:2001jy}). Reviews on Lee-Wald covariant phase space formulation can be found in \cite{Hajian:2015eha, Seraj:2016cym, Corichi:2016zac}. In the next section, we will focus on a specific method in this formulation,  the ``solution phase space method" (SPSM) introduced in Ref.\cite{Hajian:2015xlp}, which fits well to our analysis there. Interested readers can refer to \cite{Hajian:2016kxx,Ghodrati:2016vvf,Hajian:2016iyp,Hajian:2017mrf} for reviews and applications of SPSM (and \cite{Hajian:2015eha} as its precursor).

To begin, let us denote generators of diffeomorphism and gauge transformations by $\e\equiv\{\xi^\mu, \boldsymbol{\lambda}\}$, where $\xi^\mu$ corresponds to diffeomorphism, while $\boldsymbol{\lambda}$ is generator of the gauge transformations of $\boldsymbol{A}$. Under this set of transformations,  dynamical fields transform as
\begin{equation}\label{field var}
\d_\e g_{\mu\nu}=\mathscr{L}_\xi g_{\mu\nu}, \qquad\qquad  \d_\e \boldsymbol{A}=\mathscr{L}_\xi \boldsymbol{A}+d\boldsymbol{\lambda},
\end{equation}
where $\mathscr{L}_\xi$ represents Lie derivative along the vector $\xi^\mu$. In the Lee-Wald formulation, a charge variation can be attributed to an arbitrary $\epsilon$. This charge variation can be calculated by a codimension-2 surface integration over a $(D-2)$ form $\boldsymbol{k}_\epsilon$. Details of calculating $\boldsymbol{k}_\epsilon$ from a given covariant Lagrangian can be found e.g. in \cite{Ghodrati:2016vvf}. For our \textit{\textbf{E}instein} and \textit{\textbf{F}orm-field} Lagrangians in $I$, these standard computations give
\be \label{k_epsilon}
\boldsymbol{k}_\e=\frac{\sqrt{-g}\left(k^{E{} \mu\nu}_\e\!+\!k^{F{} \mu\nu}_\e\right)}{2!(D-2)!}\e_{\mu\nu\mu_1\dots \m_{D-2}}dx^{\m_1}\wedge\dots\wedge dx^{\m_{D-2}},\nonumber
\ee
where the $k^{E{} \mu\nu}_\e$ and $k^{F{} \mu\nu}_\e$ can be found to be
\begin{align}\label{k}
k^{E {}\mu\nu}_\e\!=&\frac{1}{16 \pi G}\Big[\xi^\n\nabla^\m h^\alpha_{\,\,\alpha}-\xi^\n\nabla_\r h^{\m\r}+\xi_\r\nabla^\n h^{\m \r}+\frac{1}{2}h^\alpha_{\,\,\alpha}\nabla^\n\xi^\m-h^{\r\n}\nabla_\r\xi^\m\Big]-[\mu\leftrightarrow \nu],
 \nonumber\\
k^{F{} \mu\nu}_\e=&\frac{\pm 1}{8\pi G(D-2)!}\Big[\big(\frac{-h^\alpha_{\,\,\alpha}}{2}F^{\m\n\r_3\dots \r_D}+2 h^{\m\beta} F_\beta{}^{\n\r_3\dots\r_D}-\d F^{\m\n\r_3\dots\r_D}\big) (\xi^\sigma A_{\sigma\r_3\dots \r_D}+\lambda_{\r_3\dots \r_D})\nonumber\\
&\hspace*{2cm}-F^{\m\n \r_3\dots \r_D} \xi^\sigma \d A_{\sigma \r_3\dots \r_D}+(D-2)h^{\alpha\beta}F_\a{}^{\m\n\r_4\dots\r_D}(\xi^\sigma A_{\sigma\beta\r_4\dots\r_D}+\lambda_{\beta\r_4\dots\r_D})\nonumber\\
&\hspace*{2cm}+\frac{2}{D-1} F^{\m\r_2\dots\r_D}\xi^\n \d A_{\r_2\dots\r_D}\Big]-[\mu\leftrightarrow\nu],
\end{align}
with the notation $h^{\mu\nu}={\d}g^{\mu\nu}\equiv g^{\mu\alpha}g^{\nu\beta}{\d}g_{\alpha\beta}$  and ${\d}F^{\r_1\dots\r_D}{\equiv}g^{\r_1\mu_1}\dots g^{\r_D\mu_D}{\d}F_{\mu_1\dots\mu_D}$ for the metric and field strength variations.
It is worth mentioning that in computation of a generic $\boldsymbol{k}_\e$, the Lee-Wald symplectic structure (and consequently, the $\boldsymbol{k}_\e$)  remains intact under a shift of the Lagrangian by boundary terms as in \eqref{Action_Shifted}. This justifies our freedom to choose $I$ to be either the action \eqref{Action} or \eqref{Action_Shifted}. Having the $\boldsymbol{k}_\e$ in our hands, the charge variation ${\d}H$ of the generator $\epsilon$ is found by the integration
\begin{equation}\label{del H}
\d {H}_\epsilon= \oint_{\partial\Sigma} \boldsymbol{k}_\e.
\end{equation}
In the Lee-Wald formulation, $\partial\Sigma$ is the asymptotic codimension-2 boundary of the spacelike (Cauchy) hypersurfaces $\Sigma$ (i.e time foliations) of the space-time. Moreover, the relation above for ${\d}H_\e$ is an on-shell relation. In other words, in order to find ${\d}H_\epsilon$,  a solution of the equations of motion as well as a (linearized) perturbation are needed as inputs of $\boldsymbol{k}_\e$ in \eqref{k}.

For an arbitrary $\epsilon$, conservation and integrability of ${\d}H_\e$ are not guaranteed. If ${\d}H_\e$ is independent of the choice of hypersurface $\Sigma$, it is called to be conserved. If  $\d H_\e$ is variation of a charge $H_\e$, i.e. ${\d}H_\e={\d}(H_\e)$, it is integrable. Fortunately for the $\epsilon$ which we will choose, ${\d}H_\epsilon$ is conserved and integrable, as we describe it now.

In the Lee-Wald formulation, conservation is achieved by imposing some fall-off conditions on the solutions and the perturbations (see e.g. \cite{Wald:1999wa}). Nonetheless, there is a subset of symmetries whose charges are automatically conserved. This subset is called \emph{symplectic symmetry} (crystalized in \cite{CHSS:2015mza,CHSS:2015bca}). For a symplectic symmetry generator $\e$, $\boldsymbol{k}_\e$ is closed on-shell, $d\boldsymbol{k}_\e=0$, which makes the conservation of ${\d}H_\e$ to be automatic (see e.g. \cite{Hajian:2015xlp,Ghodrati:2016vvf} as reviews). Moreover owing to the Stokes' theorem, $\partial\Sigma$ in \eqref{del H} can be relaxed from the boundary into the bulk $\Sigma$, denoted by a closed smooth codimension-2 surface $\mathscr{S}$. Notice that because of topological obstructions, $d\boldsymbol{k}_\e=0$ via Stokes' theorem does not make the charge variation in \eqref{del H} to vanish trivially. One family of symplectic symmetries which we focus on here is the family of \emph{exact} symmetries. An exact symmetry generator is an $\epsilon$ which does not transform the dynamical fields at all \cite{Barnich:2003xg,Hajian:2015xlp}, i.e equations in \eqref{field var} vanish for such an $\epsilon$. For clarity, let us denote such an exact symmetry generator by $\eta$.  Here, we single out one of the exact symmetries; ``global part" of the gauge symmetry.
\begin{center}
\textit{Considering the set of closed gauge transformation generators $d\boldsymbol{\lambda}=0$, global part of the gauge transformation is defined to be $\eta_{_C}\equiv\{0,\hat{\boldsymbol{\lambda}}\}$ in which $\hat{\boldsymbol\lambda}$ is the $\boldsymbol{\lambda}$ normalized by the non-zero constant $\oint_{\mathscr{S}}\boldsymbol{\lambda}$.}
\end{center}
Notice that one can always find such $\boldsymbol{\lambda}$'s; choosing any surface $\mathscr{S}$ parametrized by some coordinates $(y^1,\dots,y^{D-2})$, the ($D-2$)-form $\boldsymbol{\lambda}=dy^1\wedge\dots\wedge dy^{D-2}$ fulfills the conditions in the definition. Note also that the normalization (and closed-ness) of $\hat{\boldsymbol{\lambda}}$ is conceptually the same as the normalization of the global part of gauge transformation in the Maxwell electrodynamics which generates the electric charge. Nonetheless, here it needs more clarifications. Specifically, it is not obvious why the normalization factor $\oint_{\mathscr{S}}\boldsymbol{\lambda}$ is a constant;  Notice that by $d\boldsymbol{\lambda}=0$ and Stokes' theorem, $\oint_{\mathscr{S}}\boldsymbol{\lambda}$ is  independent of the choice of $\mathscr{S}$ on a given $\Sigma$. However, one can show that it does not depend on the choice of $\Sigma$ too. To this end, we note that in the analysis below if we calculate ${\d}H$ for $\{0,\boldsymbol{\lambda}\}$ instead of $\{0,\hat{\boldsymbol{\lambda}}\}$, we reach to the first equation in \eqref{H_eta}, where $\hat{\boldsymbol{\lambda}}\to{\boldsymbol{\lambda}}$. It shows that $\oint_{\mathscr{S}}\boldsymbol{\lambda}$ is the charge of an exact symmetry $\{0,\boldsymbol{\lambda}\}$, which is automatically conserved \cite{Hajian:2015xlp}, and hence independent of $\Sigma$.  As a result, this ($D-2$) dimensional integration which is independent of $\mathscr S$ \emph{and} $\Sigma$ (i.e. independent of two coordinates which are not integrated over) yields a constant result. By this analysis, we deduce that $d\hat{\boldsymbol{\lambda}}=0$, implying that $\eta_{_C}$ in the definition above is an exact symmetry $\d_{\eta_{_C}}\! \boldsymbol{A}\!=\!0$. 

Now, we are ready to calculate the conserved charge associated with the global gauge symmetry. To find ${\d}C\equiv{\d}H_{\eta_{_C}}$, one should replace $\epsilon$ in \eqref{k} by $\{0,\hat{\boldsymbol{\lambda}}\}$. The result simplifies to variation of the standard Noether charge
\begin{align}\label{Standard Noether}
\d C=\d \oint_{\mathscr{S}}\mathbf{Q}_{\eta_{_C}} \quad \Rightarrow \quad C=\oint_{\mathscr{S}}\mathbf{Q}_{\eta_{_C}},
\end{align}
where $\mathbf{Q}_{\eta_{_C}}=\star\mr{Q}_{\eta_{_C}}$ and
\begin{align}
\mr{Q}_{\eta_{_C}}^{\mu\nu}=\frac{\mp 1}{4\pi G(D-2)!}F^{\mu\nu\r_3\dots\r_D}\hat{\lambda}_{\r_3\dots\r_D}.
\end{align}
Inserting the on-shell $F^{\mu\nu\r_3\dots\r_D}$ from the solution \eqref{F_Solution},
\begin{align}\label{H_eta}
C=\frac{\pm c}{4{\pi}G}\oint_{\mathscr{S}}\hat{\boldsymbol{\lambda}}=\frac{\pm c}{4{\pi}G}.
\end{align}
Recalling $\Lambda=\pm c^2$ from \eqref{Lambda_c dynamic} or \eqref{ActionCosmological-}, then
\begin{equation}\label{main}
\boxed{C=\frac{\pm \sqrt{|\Lambda|}}{4\pi G},\quad \text{(or)}\quad \Lambda=\pm(4\pi G)^2 C^2,}
\end{equation}
which is our main result in this paper.

$\Lambda$ in \eqref{main} is conserved irrespective of dynamics of the metric or the gauge field. This can have implications for the \emph{dynamics} of the universe, e.g. protection of $\Lambda$ by the gauge symmetry. In addition, promoting $\Lambda$ to a conserved charge can lead to new insights on \emph{thermodynamics} of the gravitational systems. In this latter direction, we will show how $\Lambda$ enters the first law of black hole thermodynamics in the next section.

\section{Cosmological conserved charge in the first law of thermodynamics}

Regarding $\Lambda$ as a conserved charge, $\d\Lambda$ is expected to contribute to the first law naturally (in contrast with its orthodox role as a parameter in the Lagrangian). Here we propose a chemical potential conjugate to $\Lambda$ which is defined independently of the first law. So, the $\d\Lambda$ term in the first law would be a non-trivial contribution.

To proceed, we invoke the SPSM which was advertised in the previous section. After introducing the chemical potential $\mathit{\Theta}$, we reconsider the entropy as a conserved charge \cite{Wald:1993nt,Iyer:1994ys}. Then \emph{based on these new proposals}, we prove the first law of black hole thermodynamics in the presence of $\Lambda$. In the end, we show how  $\mathit{\Theta}{\d}C$ term is related to the pressure term $V\d P$ in the literature.

SPSM is a specification/relaxation of the Lee-Wald covariant phase space formulation in different aspects. Here, we mention two main aspects which are related to our analysis. The first aspect is concentrating on the symplectic symmetries instead of arbitrary $\epsilon$. As mentioned, this specification makes ${\d}H_\e$ to be automatically conserved. Besides, integrations can be performed on arbitrary $\mathscr{S}$, closed codimension-2 surfaces smoothly deformed from $\partial\Sigma$ into the bulk. In addition, focusing on exact symplectic symmetries removes the ambiguities in the definition of Lee-Wald charges. The second aspect in SPSM is ``generalizing" the generator of the entropy as a conserved charge. In \cite{Wald:1993nt,Iyer:1994ys}, Iyer and Wald showed that entropy for non-extremal black holes is conserved charge of the horizon Killing vector $\zeta_{_\mr{H}}^\mu$ normalized by the Hawking temperature $T_{_{\mr H}}=\frac{\kappa_{_\mr{H}}}{2\pi}$, where $\kappa_{_{\mr{H}}}$ is the surface gravity of the Killing horizon H. Later in \cite{Hajian:2013lna}, infinite number of horizon Killing vectors whose charges are the entropy of \emph{extremal} black holes were found in the region near extremal horizons. Nonetheless, in the presence of electromagnetic gauge fields (with the gauge symmetry $A_\mu\to A_\mu+\partial_\mu\lambda$), integrability shows that the proposed Killing vectors (both for extremal and non-extremals) have missed a contribution from the gauge fields \cite{Hajian:2015xlp}. Ameliorating this in SPSM, the ``generalized" exact symmetry generator $\epsilon=\{\xi^\mu,\lambda\}$ for the entropy is
\begin{equation}\label{eta S EM}
\eta_{_S}=\frac{2\pi}{\kappa_{_{\mr{H}}}}\{\zeta^\mu_{_{\mr{H}}},-\mathit{\Phi}_{_{\mr{H}}}\}, \qquad\qquad \mathit{\Phi}_{_{\mr{H}}}=\zeta_{_{\mr{H}}}\cdot A\big|_{\mr{H}}.
\end{equation}
In words, the Iyer-Wald generator $\frac{2\pi\zeta_{_{\mr{H}}}}{\kappa_{_{\mr{H}}}}$ is accompanied with the global gauge transformation $\lambda=1$ multiplied by the factor $\frac{-2\pi\mathit{\Phi}_{_{\mr{H}}}}{\kappa_{_{\mr{H}}}}$, where $\mathit{\Phi}_{_{\mr{H}}}$ is electric potential of the electromagnetic gauge field $A_\mu$ on the horizon. We note that $\eta_{_S}$ is an exact symmetry. So in SPSM, entropy $S=H_{\eta_{_S}}$ can be calculated on any surface $\mathscr{S}$. This simplifies the proof of the first law, as we will show in a moment.

Having in mind that electric charge is the charge of the electromagnetic global gauge symmetry and $\mathit{\Phi}_{_{\mr H}}$ is its conjugate chemical potential, we return to our action $I$, which has the gauge field $\boldsymbol{A}$ instead of the electromagnetic field $A$. The $C$ in \eqref{H_eta} is the charge of the global part of the gauge symmetry analogous to the electric charge. Using this analogy, motivated by the electric potential in \eqref{eta S EM} one can have the following definition.
\begin{center}
\textit{Conjugate chemical potential of $C$ can be defined to be}
\begin{equation}\label{Theta}
\boxed{\mathit{\Theta}_{_\mr{H}}\equiv\oint _{\mr{H}}\zeta_{_{\mr{H}}}\cdot \boldsymbol{A}.}
\end{equation}
\end{center}
In addition, generator of the entropy in this context is generalized to have a new term proportional to $\hat{\boldsymbol{\lambda}}$
\begin{equation}\label{eta S}
\eta_{_S}=\frac{2\pi}{\kappa_{_{\mr{H}}}}\{\zeta^\mu_{_{\mr{H}}},-\mathit{\Phi}_{_{\mr{H}}},-\mathit{\Theta}_{_{\mr{H}}}\hat{\boldsymbol{\lambda}}\}.
\end{equation}
It is obvious that by keeping $-\mathit{\Phi}_{_{\mr{H}}}$, we have considered the presence of electromagnetic field in the theory too.\\

\noindent\textbf{Proof of the first law}\\
Focusing on symplectic symmetries, the first law of black hole thermodynamics is directly derived from the local identity relating the generator of charges. Let us denote the symplectic symmetry generators of the mass $M$, angular momenta $J_i$ (for some number $i$) and electric charge $Q$ by $\eta_{_M}=\{\partial_t,0,0\}$, $\eta_{_{J_i}}=\{-\partial_{\varphi^i},0,0\}$ and $\eta_{_Q}=\{0,1,0\}$ respectively, written in appropriate coordinates. Denoting horizon angular velocities by $\mathit{\Omega}^i_{_\mr{H}}$, and by $\zeta_{_{\mr{H}}}=\partial_t+\mathit{\Omega}^i_{_\mr{H}}\partial_{\varphi^i}$ in \eqref{eta S}, the relation below follows.
\begin{align}\label{eta local constraint}
\frac{\kappa_{_\mr{H}}}{2\pi}\eta_{_S}= \eta_{_M}-\mathit{\Omega}^i_{_\mr{H}} \eta_{_{J_i}}-\mathit{\Phi}_{_\mr{H}}\eta_{ _Q}-\mathit{\Theta}_{_\mr{H}} {\eta_{_C}}.
\end{align}
Then, by the linearity of  charge variations ${\d}H_{a\epsilon_1+b\epsilon_2}=a{\d}H_{\epsilon_1}+b\,{\d}H_{\epsilon_2}$ in \eqref{del H}, the first law of (black hole) thermodynamics is derived from \eqref{eta local constraint}. This generalized first law has a new term coming  from the gauge field $\boldsymbol{A}$
\begin{equation}\label{first law}
\boxed{T_{_\mr{H}}\d S= \d M - \mathit{\Omega}^i_{_\mr{H}}\d J_i -\mathit{\Phi}_{_\mr{H}} \d Q-\mathit{\Theta}_{_\mr{H}}\d C.}
\end{equation}\\

\noindent\textbf{Example: Kerr-(A)dS black hole}\,\\
Asymptotically non-rotating Kerr-(A)dS black hole is
\begin{align}\label{Kerr AdS metric}
ds^2=& -\Delta_\theta(\frac{1-\frac{\Lambda r^2}{3}}{\Xi}-\Delta_\theta f)dt^2+\frac{\rho ^2}{\Delta_r}dr^2+\frac{\rho ^2}{\Delta_\theta} d\theta ^2
-2\Delta_\theta fa\sin ^2 \theta\,dt d\varphi\nonumber \\
& +\big( \frac{r^2+a^2}{\Xi}+fa^2\sin ^2\theta \big)\sin ^2\theta\,d\varphi ^2,\qquad \qquad \rho^2 \equiv r^2+a^2 \cos^2 \theta\,,\nonumber \\
& \hspace*{-0.9cm}\Delta_r \equiv (r^2+a^2)(1-\frac{\Lambda r^2}{3})-2Gmr,\quad
\Delta_\theta\equiv 1+\frac{\Lambda a^2}{3}\cos ^2\theta\,,\qquad \Xi\equiv 1+\frac{\Lambda a^2}{3}\,,\quad
f\equiv\frac{2Gmr}{\rho ^2\Xi^2}.\nonumber
\end{align}
 ``Appropriately gauge fixed" $\boldsymbol{A}$ field for this geometry can be found to be
\begin{equation}\label{Kerr AdS A}
\hspace*{-0.1cm}\boldsymbol{A}\!=\!-\frac{\sqrt{|\Lambda|}(r^3+3ra^2\cos ^2\theta \!+\!\frac{Gma^2}{\Xi})\sin\theta}{3\Xi}dt\wedge d\theta \wedge d\varphi.
\end{equation}
This black hole is a solution to the equations of motion \eqref{EH equation}, and has three solution parameters $m$, $a$ and $\Lambda$. Chemical potentials for this solution can be calculated as
\begin{align}
\kappa_{_\mr H}=\frac{r_{_\mr H}(1-\frac{\Lambda a^2}{3}-\Lambda{r_{_\mr H}^2}-\frac{a^2}{r_{_\mr H}^2})}{2(r_{_\mr H}^2+a^2)},\quad \mathit{\Omega}_{_\mr H}=\frac{a(1-\frac{\Lambda r_{_\mr H}^2}{3})}{r_{_\mr H}^2+a^2},\quad \mathit{\Theta}_{_\mr{H}}=-\frac{\sqrt{|\Lambda|}\,\,4\pi (r_{_\mr H}^3+r_{_\mr H} a^2 +\frac{Gma^2}{\Xi})}{3\Xi},\label{Kerr AdS chem}
\end{align}
in which $r_{_{\mr{H}}}$ is the radius of inner, outer or cosmological horizons. Inserting the generators $\eta$ of \eqref{eta local constraint} into the $\boldsymbol{k}_\epsilon$ \eqref{k_epsilon}, conserved charges can be found by SPSM  without ambiguity and regularizations (see \cite{Hajian:2016kxx} for the details)
\begin{equation}
\hspace*{-0.3cm}M=\!\frac{m}{\Xi^2},\qquad J=\frac{ma}{\Xi^2},\qquad  S=\!\frac{\pi(r_{_\mr{H}}^2+a^2)}{G\Xi}, \qquad C=\!\frac{\pm\sqrt{|\Lambda|}}{4\pi G},
\end{equation}
satisfying the first law $T_{_\mr{H}}\d S=\d M-\mathit{\Omega}_{_\mr{H}}\d J-\mathit{\Theta}_{_\mr{H}}{\d}C$. The upper/lower signs are for Kerr-dS/Kerr-AdS black holes.\\

\noindent\textbf{Discussion I: ${{\boldsymbol{\mathit{\Theta}}}}_{_\mathbf{H}}\boldsymbol{\d}\boldsymbol{C}$ vs. $\boldsymbol{V{\d}P}$}\\
As mentioned in the introduction, the cosmological constant may be considered as a pressure term $P=-\frac{\Lambda}{8{\pi}G}$ in cosmology and the first law of thermodynamics. Hereafter, we assume that this relation is correct. The conjugate chemical potential to $P$ is called ``thermodynamic volume" denoted by $V$. As the name suggests, this quantity is usually defined/found via the first law itself or the Smarr formula (see examples in e.g. \cite{Cvetic:2010jb,Dolan:2012jh}). So, the additional $V{\d}P$ term in the first law respects the first law trivially. However in certain cases, $V$ can have geometrical meaning of the ``volume inside the black hole," usually called ``geometric volume." For clarity, we distinguish the thermodynamic and geometric volumes by the notation $V$ and $\bar V$ respectively.

At first glance, the new term $\mathit{\Theta}_{_\mr{H}}{\d}C$ in the first law \eqref{first law} does not resemble the $V{\d}P$  so much. Our last analysis in this paper is to show how these are related. By \eqref{main}
\begin{equation}
\d C=\frac{\d \Lambda}{8\pi G \sqrt{|\Lambda|}} \quad \Rightarrow \quad \d P=-\sqrt{|\Lambda|} \,\d C,
\end{equation}
which suggests that the pressure can be dealt as a conserved charge. Thermodynamic volume $V$ is defined to be the conjugate of $P$ in the first law. Therefore,
\begin{equation}
\mathit{\Theta}_{_\mr{H}}\d C=V\d P \quad \Rightarrow\quad V=-\frac{\mathit{\Theta}_{_\mr{H}}}{\sqrt{|\Lambda|}}.
\end{equation}
This result accompanied with \eqref{Theta} identifies the thermodynamic volume $V$ for the horizons of a given (black hole) solution, and re-expresses the original first law in \eqref{first law} as
\begin{align}
T_{_\mr{H}}\d S= \d M - \mathit{\Omega}^i_{_\mr{H}}\d J_i -\mathit{\Phi}_{_\mr{H}} \d Q -V\d P.
\end{align}

The last question we try to answer is the relation between $\mathit{\Theta}_{_\mr{H}}$ (and hence the $V$) and geometric volume $\bar V$. To find the answer, by Stokes' theorem we can write \eqref{Theta} as
\begin{align}\label{Theta V}
\mathit{\Theta}_{_\mr{H}}=\oint _{\mr{H}}\zeta_{_{\mr{H}}}\cdot \boldsymbol{A}=\int_{\Sigma_{\text{in}}}\!\!d(\zeta_{_{\mr{H}}}\cdot \boldsymbol{A})\,\,+\tilde{\mathit{\Theta}}_{_\mr{H}}.
\end{align}
The  term $\tilde{\mathit{\Theta}}_{_\mr{H}}$ comes from the blindness of Stokes' theorem to the addition of a closed form to the $\zeta_{_{\mr{H}}}\cdot\boldsymbol{A}$, which itself can originate from the gauge freedom of $\boldsymbol{A}$. Rephrasing in other words, $\mathit{\Theta}_{_\mr{H}}$ in \eqref{Theta} is sensitive to the choice of the gauge for $\boldsymbol{A}$, while the integrand in the integration over $\Sigma_{\text{in}}$ in \eqref{Theta V} is not (if $\mathscr{L}_{\zeta_{_{\mr{H}}}}\!\boldsymbol{A}=0$ which we assume to be satisfied). The gauge can be appropriately fixed by requiring \emph{integrability for the entropy} over solution phase space; A nice feature of SPSM is that in this method, phase space and its tangent space are determined explicitly (in contrast with phase spaces determined by some fall-off conditions). Hence, integrability condition can be checked completely (see e.g. \cite{Hajian:2016iyp} for more details), constraining the choice of gauge. For instance in the Kerr-(A)dS example, the gauge fixing term $\tilde{\mathit{\Theta}}_{_\mr{H}}$ is the last term of $\mathit{\Theta}_{_\mr{H}}$ in \eqref{Kerr AdS chem}, which contributes to the $\eta_{_S}$ in \eqref{eta S} and makes the entropy to be integrable.  Now, by the identity $d(\xi\cdot\boldsymbol{A})=\mathscr{L}_\xi \boldsymbol{A}-\xi\cdot d\boldsymbol{A}$ which is correct for any vector $\xi$, and by $\mathscr{L}_{\zeta_{_{\mr{H}}}}\!\boldsymbol{A}=0$, we have $d(\zeta_{_{\mr{H}}}\cdot\boldsymbol{A})=-\zeta_{_{\mr{H}}}\cdot\boldsymbol{F}$. Replacing this in \eqref{Theta V} and by inserting $\boldsymbol{F}$ from the \eqref{F_Solution}, we find
\begin{align}\label{V geometric}
\mathit{\Theta}_{_\mr{H}}=-\sqrt{|\Lambda|} \int_{\Sigma_{\text{in}}}\frac{\sqrt{-g}\,\zeta^\mu_{_{\mr{H}}}\epsilon_{\mu\r_2 ...\r_D}\,dx^{\r_2}\wedge\dots\wedge dx^{\r_D}}{(D-1)!}+\tilde{\mathit{\Theta}}_{_\mr{H}}.
\end{align}
The integration above can be dubbed as ``geometric volume" 
\begin{equation}
\bar V\equiv \int_{\Sigma_{\text{in}}}\frac{\sqrt{-g}\,\zeta^\mu_{_{\mr{H}}}\epsilon_{\mu\r_2 ...\r_D}\,dx^{\r_2}\wedge\dots\wedge dx^{\r_D}}{(D-1)!}.
\end{equation}
The $\Sigma_{\text{in}}$ is a hypersurface inside the horizon. Our result \eqref{V geometric} beautifully shows why $\sqrt{-g}$ (and not determinant of the induced metric on $\Sigma_{\text{in}}$) appears in definition of the geometric volume. Besides, the presence of $\tilde{\mathit{\Theta}}_{_\mr{H}}$ in \eqref{V geometric} sheds light on why the geometric and thermodynamic volumes do not always coincide \cite{Cvetic:2010jb,Dolan:2012jh} (e.g. for the Kerr-(A)dS black hole which was exemplified above).\\

\noindent\textbf{Discussion II: Pressure from a conserved charge}\\
One might ask how the pressure can be related to a conserved charge. An answer might be suggested from the similarity of the top-form formalism and electromagnetism; a box of gas of some (positively \emph{or} negatively) electrically charged molecules would indicate more pressure than the same gas of neutral molecules; more electric charge yields more pressure.\\

\section{Conclusion}
Variation of mass, angular momenta, electric charge and entropy in the first law of black hole thermodynamics are all conserved charge variations. This has been recognized by realization of the Wald (Noether charge) entropy in 1993 \cite{Wald:1993nt}. On the other hand, cosmological constant also contributes to the first law as a pressure term. We showed that it is also a conserved charge. To this end, we studied the cosmological constant coming from the on-shell value of a top-form, and crystallized the global part of the gauge symmetry as the corresponding symmetry.  This natural recognition also admits the conjugate chemical potential to this conserved charge (the $\mathit{\Theta}_{_\mr{H}}$ in equation \eqref{Theta}), which can settle down the ``volume" problem in the context of black hole thermodynamics. Our proposed chemical potential also suggested a rigorous relation for the geometric volume. Finally, we proved the first law with the pressure term in its natural context. 

Exploring the $\mathit{\Theta}_{_\mr{H}}$ for different solutions in different gravitational theories can be a good practice to examine our analysis. Besides, conservation of cosmological constant, as well as the gauge field from which it originates can change the way we think about this mysterious constant, as well as the successful $\Lambda$-CDM theory. It might be the opening window to a new insight on the important dark energy problem. Quantization of the gauge field, as well as probable protection of $\Lambda$ by the gauge symmetry can lead to interesting results in this subject. \\

\textbf{Acknowledgments:} The authors would like to thank the scientific atmosphere of Yerevan FAR/ANSEF-ICTP 2017 Summer school
in theoretical physics, where this work initiated, and KH would like to thank the ICTP network scheme NT-04.
He also thanks Shahin Sheikh-Jabbari and Erfan Esmaeili for helpful discussions. The work of DC was supported by the Tomsk Polytechnic University competitiveness enhancement program, the RF Presidential grant MK-2101.2017.2, and the RFBR grant 18-52-05002.

\end{document}